\documentstyle[aps,epsfig]{revtex}
\begin{document}

\title{Nuclear incompressibility using the density dependent M3Y effective interaction }

\author{D.N. Basu\thanks{E-mail:dnb@veccal.ernet.in}}
\address{Variable  Energy  Cyclotron  Centre,  1/AF Bidhan Nagar,
Kolkata 700 064, India}
\date{\today }
\maketitle
\begin{abstract}

      A density dependent M3Y effective nucleon-nucleon (NN) interaction which was based on the G-matrix elements of the Reid-Elliott  NN potential has been used to determine the incompressibity of infinite nuclear matter. The nuclear interaction potential obtained by folding in the density distribution functions of two interacting nuclei with this density dependent M3Y effective interaction had been shown earlier to provide excellent descriptions for medium and high energy $\alpha$ and heavy ion elastic scatterings as well as $\alpha$ and heavy cluster radioactivities. The density dependent parameters have been chosen to reproduce the saturation energy per nucleon and the saturation density of spin and isospin symmetric cold infinite nuclear matter. The result of such calculations for nuclear incompressibility using the density dependent M3Y effective interaction based on the G-matrix elements of Reid-Elliott NN potential predicts a value of about 300 MeV for nuclear incompressibility. 
     
\end{abstract}



      The compression modulus or the incompressibility of infinite nuclear matter $K_\infty$, defined through the curvature of the energy per nucleon  E/A=$\epsilon$ of the cold spin and isospin symmetric infinite nuclear matter at the saturation density 
is of fundamental importance. Different types of nuclear equation of state (EOS) considered in the study of central heavy ion collisions \cite{r1} are normally distinguished by different $K_\infty$ values. Experimentally, a widely used method is the determination of the incompressibility from the observed giant monopole resonances (GMR) \cite{r2,r3,r4}. But experimentally extracted incompressibility values \cite{r4} from the GMR data still show the span of $K_\infty$ values no better than 1.7 $\times$ (200-300) MeV, and more precise data are needed to pin down the range. Evidence from the various nuclear data and neutron star masses favor a high compression modulus $K_\infty$ (approximately equals) 300 MeV \cite{r5}. An accurate determination of $K_\infty$ is very important for the study of properties of nuclei (radii, masses, giant resonances etc.), supernova collapses, neutron stars, and heavy ion collisions.

      A well-defined effecttive nucleon-nucleon (NN) interaction in the nuclear medium is important not only for different structure models but also for the microscopic calculation of the nucleon-nucleus and nucleus-nucleus potentials used in the analysis of the nucleon- and heavy-ion scattering. Effective NN interaction can be best constructed from a sophisticated G-matrix calculation. This interaction has been derived by fitting its matrix elements in an oscillator basis to those elements of the G-matrix obtained with the Reid-Elliott soft-core NN interaction \cite{r6}. The ranges of the M3Y forces were chosen to ensure a long-range tail of the one-pion exchange potential as well as a short range repulsive part simulating the exchange of heavier mesons. Such an effective NN interaction has been shown to provide a more realistic shape of the scattering potentials of the nucleon or heavy ion optical potentials obtained by folding in the density distribution functions of two interacting nuclei with the effective NN interaction \cite{r7}.           
 
      The real part of the nuclear interaction potential obtained by folding in the density distribution functions of two interacting nuclei with the M3Y effective interaction with one nucleon exchange evaluated from first principles was shown to provide good descriptions for medium and high energy $\alpha$ and heavy ion elastic scatterings \cite{r8,r9}. It was observed that the high energy  $\alpha$ and heavy ion elastic scattering data were better fitted, requiring less renormalisation and better chi-square per degrees of freedom ($\chi^2/F$), when folding model potentials obtained from M3Y effective interaction based on the G-matrix elements of Reid-Elliott NN potential were used instead of those obtained from M3Y effective interaction based on the G-matrix elements of Paris NN potential. The semirealistic explicit density dependence on the M3Y effective interaction with a zero-range pseudo-potential was then employed. The zero-range pseudo-potential represents the single-nucleon exchange term while the density dependence takes care off the higher order exchange effects and the Pauli blocking effects. Since the density dependence of the effective projectile-nucleon interaction has been found to be fairly independent of the projectile, as long as the projectile-nucleus interaction is amenable to a single-folding prescription, implies that in a double folding model, the density dependent effects on the nucleon-nucleon interaction can be factorized into a target term times a projectile term \cite{r10,r11}. The folding model potentials obtained using such factorized density dependent M3Y-Reid-Elliott effective interaction along with a zero-range pseudo-potential were then used to describe successfully the medium and high energy $\alpha$ and heavy ion elastic scatterings \cite{r12,r13}. 

      Subsequently, the folding model potentials obtained using M3Y effective interaction based on the G-matrix elements of Reid-Elliott NN potential were then used to describe successfully the heavy particle decay \cite{r14} and $\alpha$ decay \cite{r15} while it was found to provide excellent description for cluster radioactivity when the spin-parity conservation was also taken into account \cite{r16}. In the present work nuclear incompressibility has been calculated theoretically using the same density dependent M3Y effective interaction, based on the G-matrix elements of Reid-Elliott NN potential, supplemented by a zero range pseudo-potential. It is worthwhile to mention here that due to attractive character of the M3Y forces the saturation condition for cold nuclear matter is not fulfilled. However, the realistic description of nuclear matter properties can be obtained with the density dependent M3Y effective interaction. Therefore, the density dependent parameters have been obtained by reproducing the saturation energy per nucleon and the saturation nucleonic density of the spin and isospin symmetric cold  infinite nuclear matter. 

      The incompressibility $K_\infty$ of the spin and isospin symmetric cold infinite nuclear matter is defined as     
\begin{equation}
 K_\infty = k_F^2\partial^2\epsilon/\partial{k_F^2} = 9\rho^2\partial^2\epsilon/\partial\rho^2\mid_{\rho=\rho_\infty}
\label{seqn1}
\end{equation}
\noindent
where the Fermi momentum $k_F$ for the spin and isospin symmetric infinite nuclear matter is given by

\begin{equation}
 k_F^3 = 1.5\pi^2\rho
\label{seqn2}
\end{equation}                                                                                                                                           
\noindent     
and $\rho$ is the nucleonic density while $\rho_{\infty}$ being the saturation density for the spin and isospin symmetric cold infinite nuclear matter. 

      The effective nucleon-nucleon interaction $v(s)$ is assumed to be density and energy dependent and therefore becomes functions of density and energy and is generally written as \cite{r12,r13} 

\begin{equation}
  v(s,\rho, \epsilon) = t^{M3Y}(s, \epsilon) g(\rho, \epsilon)
\label{seqn3}
\end{equation}   
\noindent
where $t^{M3Y}$ is given by \cite{r6} 

\begin{equation}
  t^{M3Y}(s, \epsilon) = 7999 \exp( - 4s) / (4s) - 2134 \exp( - 2.5s) / (2.5s) + J_{00}(\epsilon) \delta(s)
\label{seqn4}
\end{equation}   
\noindent
where the zero-range pseudo-potential representing the single-nucleon exchange term is given by \cite{r17} 

\begin{equation}
 J_{00}(\epsilon) = -276 (1 - 0.005\epsilon) (MeV.fm^3)
\label{seqn5}
\end{equation}   
\noindent
and the density dependent part has been taken to be \cite{r18}

\begin{equation}
 g(\rho, \epsilon) = C (1 - \beta(\epsilon)\rho^{2/3}) 
\label{seqn6}
\end{equation}   
\noindent
which takes care of the higher order exchange effects and the Pauli blocking effects.

      The energy per nucleon $\epsilon$ obtained using the effective nucleon-nucleon interaction $v(s)$ for the spin and isospin symmetric cold infinite nuclear matter is given by

\begin{equation}
 \epsilon = 3\hbar^2k_F^2/10m + g(\rho, \epsilon)\rho J_v / 2
\label{seqn7}
\end{equation}   
\noindent
where m is the nucleonic mass equal to 931.4943 $MeV/c^2$ and $J_v$ represents the volume integral of the M3Y interaction supplemented by the zero-range pseudopotential having the form    

\begin{equation}
 J_v(\epsilon)  =   \int \int \int t^{M3Y}(s, \epsilon) d^3s = 7999 (4\pi/4^3) - 2134 (4\pi/2.5^3) + J_{00}(\epsilon) 
\label{seqn8}
\end{equation}
\noindent
The equilibrium density of the nuclear matter is determined from the saturation condition $\partial\epsilon/\partial\rho = 0$ which yields the equation  

\begin{equation}
 \partial\epsilon/\partial\rho = \hbar^2k_F^2/5m\rho + J_v C (1 - 5\beta(E)\rho^{2/3}/3) /2 = 0
\label{seqn9}
\end{equation}
\noindent
while the eqn.(7) can be rewritten with the help of eqn.(6) as 

\begin{equation}
 3\hbar^2k_F^2/10m + \rho J_v C (1 - \beta(E)\rho^{2/3})/2  = \epsilon
\label{seqn10}
\end{equation}
\noindent
so that eqn.(9) and eqn.(10) can be solved simultaneously for any values of the saturation energy per nucleon and the saturation density of the spin and isospin symmetric cold  infinite nuclear matter to obtain the values of the density dependent parameters $\beta(\epsilon)$ and C. Density dependent parameters $\beta(\epsilon)$ and C, thus obtained, can be given by  

\begin{equation}
 \beta(\epsilon) = [(3-3p)/(9-5p)]/\rho^{2/3}
\label{seqn11}
\end{equation} 
\noindent
where

\begin{equation}
 p = [10m\epsilon]/[\hbar^2(1.5\pi^2\rho)^{2/3}]
\label{seqn12}
\end{equation} 
\noindent
and 

\begin{equation}
 C = -[2\hbar^2k_F^2] / [5mJ_v\rho(1 - 5\beta(\epsilon)\rho^{2/3}/3)]
\label{seqn13}
\end{equation} 
\noindent
respectively.

      Once the density dependent parameters $\beta(\epsilon)$ and C have been fixed for a saturation energy per nucleon and a saturation density of a spin and isospin symmetric cold infinite nuclear matter, the incompressibility $K_{\infty}$ can be evaluated from the equation given by  

\begin{equation}
 K_{\infty} = [-3\hbar^2k_F^2/5m - 5 J_v C \beta(\epsilon)\rho^{5/3}]_{\rho=\rho_{\infty}}
\label{seqn14}
\end{equation} 
\noindent
which has been obtained using eqn.(1), eqn.(2) and eqn.(9).

      The results of the present calculations using the saturation energy per nucleon $\epsilon$ = -16 MeV for the spin and isospin symmetric cold infinite nuclear matter have been presented in Table-1. Since the value used for the saturation density by different groups do differ, a narrow range of its acceptable values have been used. The value of the density dependence parameter $\beta(\epsilon)$, which is supposed to dependend on energy, has been found rather close to 1.6 $fm^2$, a value which was obtained by optimum fit to the experimental data for $\alpha$ radioactivity which involves very low energies \cite{r16} compared to the high energy $\alpha$ elastic scattering \cite{r12}. Using the usual value of 0.005/MeV for the parameter of energy dependence of the zero range pseudo-potential, as one can see in eqn.(5), the value obtained for the other density dependence parameter C is close to 2. However, readjusting the value of the energy dependence parameter to 0.100/MeV brings down the value of the density dependence parameter C approximately equal to 1, while the value of the nuclear incomressibility remains unaltered which is obvious from eqn.(13) and eqn.(14). The values obtained for the nuclear incompressibility ranges from 304 MeV - 310 MeV for acceptable range of values of saturation densities used here.         
 
\begin{table}
\caption{Compression modulus at different saturation densities}
\begin{tabular}{ccccc}
$\rho_{\infty}$&$\beta(\epsilon)$&C    &Energy dependence parameter& $K_{\infty}$      \\
$fm^{-3}$&$fm^2$               &      &             $MeV^{-1}$             &      $MeV$       \\ \hline

  0.170 &1.551 & 1.98 &0.005& 309.6 \\ 
  0.170 &1.551 & 1.02 &0.100& 309.6 \\ 
  0.165 &1.586 & 2.02 &0.005& 308.2 \\ 
  0.165 &1.586 & 1.04 &0.100& 308.2 \\ 
  0.160 &1.624 & 2.07 &0.005& 306.9 \\ 
  0.160 &1.624 & 1.06 &0.100& 306.9 \\ 
  0.155 &1.664 & 2.11 &0.005& 305.5 \\ 
  0.155 &1.664 & 1.09 &0.100& 305.5 \\ 
  0.150 &1.705 & 2.16 &0.005& 304.0 \\ 
  0.150 &1.705 & 1.11 &0.100& 304.0 \\ 

\end{tabular} 
\end{table}

      Theoretical estimate of $K_{\infty}$ from refractive $\alpha$-nucleus scattering claims to have pinned down its range to about 240 MeV-270 MeV \cite{r19,r20} ruling out the lower value of about 180 MeV. Another recent theoretical estimate by infinite nuclear matter model (INM) \cite{r21} claims a well defined and stable value of $K_{\infty}=288\pm20$ MeV. The finite nuclear incompressibility $K_A$ can be related to the energies $E_{GMR}$ of the giant monopole resonance (GMR) as  

\begin{equation}
 K_A = M<r^2>E^2_{GMR}/\hbar^2 
\label{seqn15}
\end{equation} 
\noindent
where M is the mass of the nucleus, $<r^2>$ is its mean square radius. Following a liquid drop model type expansion, one writes down a similar expression for $K_A$ as 
\begin{equation}
 K_A = K_v + K_s A^{-1/3} + K_{cv} A^{-2/3} + K_{asym} (1-2Z/A)^2 + K_cZ^2 A^{-4/3}
\label{seqn16}
\end{equation} 
\noindent                                 
and then the volume term $K_v$ can be identified as $K_{\infty}$. Although eqn.(16) looks to be a natural path to obtain $K_{\infty}$ from $K_A$, this expansion is beset with many difficulties in regard to its convergence and the number of data on GMR being limited, no precise value of $K_{\infty}$ can be extracted \cite{r4}. A recent determination of $K_{\infty}$ based upon the production of hard photons on heavy ion collision led to the experimental estimate $K_{\infty}=290\pm50$ MeV \cite{r22} and the result of the present calculation for $K_{\infty}$ is in excellent agreement with this recent experimental finding.         

      A density dependent M3Y-Reid-Elliott effective nucleon-nucleon interaction which has been used within a folding model prescription to provide excellent description for medium and high energy $\alpha$ and heavy ion scatterings, $\alpha$ and heavy ion cluster radioactivities, has been used to calculate the nuclear incompressibility $K_{\infty}$ of the spin and isospin symmetric cold infinite nuclear matter. The resulting value of $K_{\infty}$ is found to be 304 MeV-310 MeV approximately. This value is in excellent agreement with the recent experimental estimate of $K_{\infty}=290\pm50$ MeV and rules out lower values of about 180 MeV. Results of the present calculations are very close to the recent theoretical estimate by INM of $288\pm20$ MeV.

\end{document}